\documentclass[sigconf,nonacm]{acmart}
\usepackage{popets}

\setcopyright{popets}
\copyrightyear{YYYY}

\acmYear{YYYY}
\acmVolume{YYYY}
\acmNumber{X}
\acmDOI{XXXXXXX.XXXXXXX}
\acmISBN{}
\acmConference{Proceedings on Privacy Enhancing Technologies}
\usepackage{algorithmic}
\usepackage{graphicx}
\usepackage{stfloats}
\usepackage{textcomp}
\usepackage{xcolor}
\usepackage{graphicx}
\usepackage{caption}
\usepackage{subcaption}  
\def\BibTeX{{\rm B\kern-.05em{\sc i\kern-.025em b}\kern-.08em
    T\kern-.1667em\lower.7ex\hbox{E}\kern-.125emX}}

\usepackage[normalem]{ulem}
\usepackage{rotating}
\usepackage{tablefootnote}
\usepackage{tabularray}
\usepackage{hyperref}
\usepackage{microtype}
\usepackage{graphicx}
\usepackage{subcaption} 
\usepackage{booktabs} 

\usepackage{hyperref}

\usepackage{mathtools}
\usepackage{amsthm}

\usepackage[capitalize,noabbrev]{cleveref}

\theoremstyle{plain}
\newtheorem{theorem}{Theorem}[section]

\newtheorem{lemma}[theorem]{Lemma}

\theoremstyle{definition}
\newtheorem{definition}[theorem]{Definition}

\theoremstyle{remark}

\usepackage[textsize=tiny]{todonotes}

\usepackage{colortbl} 

\usepackage{multirow}
\newcommand{\mysplit}[1]{%
  \begin{tabular}{@{}c@{}}   
    #1
  \end{tabular}
  }
\usepackage{algorithm}      
\usepackage{caption}
\usepackage{subcaption}
\usepackage{graphicx}
\usepackage{caption}
\usepackage{subcaption}
\usepackage{caption}
\usepackage{subcaption}

\begin{document}

\title[Quadratic Gradient: Accelerating Privacy-Preserving Logistic Regression Training with Homomorphic Encryption]{Quadratic Gradient: Accelerating Privacy-Preserving Logistic Regression Training with Homomorphic Encryption}

\author{John Chiang}
\thanks{This work has been submitted to the IEEE for possible publication. Copyright may be transferred without notice, after which this version may no longer be accessible.}
\orcid{0000-0003-0378-0607}
\email{john.chiang.smith@gmail.com}

\renewcommand{\shortauthors}{John Chiang}

\begin{abstract}
Training logistic regression over encrypted data has emerged as a prominent approach to addressing security concerns in recent years. In this paper, we introduce an efficient gradient variant, termed the \textit{quadratic gradient}, which is specifically designed for privacy-preserving logistic regression while remaining equally effective in plaintext optimization. By incorporating this quadratic gradient, we enhance Nesterov's Accelerated Gradient (NAG), Adaptive Gradient (AdaGrad), and Adam algorithms. We evaluate these enhanced algorithms across various datasets, with experimental results demonstrating state-of-the-art convergence rates that significantly outperform traditional first-order gradient methods. Furthermore, we apply the enhanced NAG method to implement homomorphic logistic regression training, achieving comparable performance within only four iterations. The proposed quadratic-gradient approach offers a unified framework that synergizes the advantages of first-order gradient methods and second-order Newton-type methods, suggesting broad applicability to diverse numerical optimization tasks.
\end{abstract}


\maketitle

\section{Introduction}

\subsection{Background}
Given a patient's longitudinal healthcare data, one can train a logistic regression (LR) model to predict the likelihood of disease onset. However, personal health information is inherently sensitive, and privacy concerns remain a significant barrier to the large-scale sharing and aggregation of biomedical data. To address this, Homomorphic Encryption (HE) offers a robust security paradigm: data can be transformed into ciphertexts and outsourced to the cloud for computation without ever being decrypted, thereby preventing unauthorized access by the service provider.

The annual iDASH competition has become a benchmark for evaluating cryptographic technologies in biomedical settings. Since 2014, iDASH has promoted the development of privacy-preserving solutions for genomic and healthcare data analysis. In particular, the 2017 (Track 3) and 2018 (Track 2) competitions focused on efficient homomorphic-encryption-based training of logistic regression models, highlighting the practical importance of scalable encrypted learning techniques.

Despite substantial progress, encrypted LR training remains computationally expensive because the nonlinear sigmoid function must be approximated by low-degree polynomials, and iterative gradient-based optimization requires a large number of costly homomorphic operations. To address this limitation, we propose an enhanced quadratic-gradient training approach that achieves faster convergence while preserving prediction accuracy. By reducing the number of training iterations required to reach a target model quality, our method significantly lowers the homomorphic computation cost and improves the practicality of privacy-preserving logistic regression training.

\subsection{Related Work}
Several studies have explored the implementation of logistic regression models within homomorphic encryption  frameworks. 
Aono et al. \cite{aono2016scalable, aono2016privacy} proposed one of the earliest homomorphic encryption-based frameworks for privacy-preserving logistic regression over distributed datasets, demonstrating the feasibility of secure collaborative model training.
Kim et~al. \cite{kim2018secure} addressed the challenges of encrypted LR training by employing full-batch gradient descent and utilizing least-squares approximation for the sigmoid function. During the 2017 iDASH competition, this problem was further investigated by Bonte and Vercauteren \cite{IDASH2018bonte}, Kim et~al. \cite{IDASH2018Andrey}, Chen et~al. \cite{IDASH2018chen}, and Crawford et~al. \cite{IDASH2018gentry}. In the subsequent 2018 iDASH competition, Kim et~al. \cite{IDASH2019kim} and Blatt et~al. \cite{IDASH2019blatt} introduced efficient ciphertext packing techniques and semi-parallel algorithms to enhance computational throughput. 

Bergamaschi et al. \cite{bergamaschi2019homomorphic} leveraged the SIMD capabilities of the CKKS homomorphic encryption scheme to enable the parallel training of thousands of logistic regression models for privacy-preserving genome-wide association studies.
Kim et al. \cite{kim2019secure} combined homomorphic encryption with differential privacy to enable secure logistic regression training over horizontally distributed datasets, addressing both computation privacy and output privacy.
De Cock et al. \cite{de2019fast} developed an optimized secure multiparty computation framework for privacy-preserving logistic regression training on large-scale genomic datasets, demonstrating the feasibility of secure learning with billions of secure multiplications.
Crockett \cite{crockett2020low} optimized the homomorphic circuit for logistic regression training by reducing the multiplicative depth of each training iteration, thereby improving the efficiency of encrypted model training.
Li and Huang \cite{li2020faster} introduced a distributed HE-based training framework that accelerates encrypted logistic regression and other data mining algorithms by reducing homomorphic circuit depth through distributed computation.
Carpov et al. \cite{carpov2020privacy} developed a privacy-preserving framework for semi-parallel logistic regression training and feature selection over encrypted genomic data by leveraging the hybrid Chimera framework combining TFHE and HEAAN.
Han et al. \cite{han2020efficient} developed a hybrid privacy-preserving logistic regression framework that combines MPC and homomorphic encryption to accelerate encrypted matrix operations and reduce the computational overhead of large-scale model training and inference.
De Cock et al. \cite{de2021high} proposed a high-performance secure multi-party computation protocol for privacy-preserving logistic regression training, introducing an efficient activation function protocol and several cryptographic optimizations for large-scale genomic data.
Byun et al. \cite{byun2021parameter} introduced a parameter-free HE-friendly logistic regression framework that avoids sigmoid approximation and hyperparameter tuning through a ridge regression reformulation.
Hong et al. \cite{hong2024privacy} developed a homomorphic encryption-based privacy-preserving inference framework for logistic and linear regression on genomic data, addressing model confidentiality by encrypting both patient data and model parameters.
Montero et al. \cite{montero2024machine} developed a TFHE-based encrypted training framework that supports privacy-preserving logistic regression and neural network training over confidential datasets.

While these works have explored various aspects of encrypted optimization, most of them rely on first-order gradient-based algorithms, such as Nesterov Accelerated Gradient (NAG), which require careful tuning of learning rate parameters. Such parameter selection is particularly challenging in the encrypted domain due to the difficulty of performing validation on ciphertexts. Bonte and Vercauteren \cite{IDASH2018bonte} proposed a practical second-order optimization method, known as the Simplified Fixed Hessian (SFH) method, which eliminates the need for learning rate tuning. However, SFH cannot be directly applied to sparse datasets, such as MNIST. Ogilvie et al. \cite{ogilvie2020improved} further extended the SFH method to CKKS-based logistic and ridge regression training by exploiting SIMD packing. Despite this, they did not recognize the inherent limitations of SFH on sparse datasets and still relied on data normalization to a non-negative range. In this work, we reveal and address these limitations of SFH. Furthermore, we extend the SFH approach and incorporate the ciphertext packing mechanism proposed by Kim et al.~\cite{IDASH2018Andrey} to achieve efficient homomorphic computation.

\subsection{Contributions}
Our specific contributions in this paper are summarized as follows:

\begin{enumerate}
    \item We introduce a novel gradient variant, termed the \textit{quadratic gradient}, which bridges first-order gradient-based methods and the second-order Newton-type algorithms. This enables a unified framework that synergistically leverages the computational efficiency of first-order methods and the rapid convergence of second-order approaches.

    \item We develop three enhanced optimization algorithms by integrating the quadratic gradient into existing frameworks. Experimental results demonstrate that these enhanced algorithms achieve state-of-the-art performance in terms of convergence rates across diverse datasets.

    \item We implement privacy-preserving logistic regression training using the enhanced Nesterov’s Accelerated Gradient (NAG) method. To the best of our knowledge, our implementation strikes an optimal balance between computational efficiency and storage overhead without significant performance degradation.

    \item We propose a systematic framework that formalizes the derivation of constant Hessian approximations for general optimization objectives, thereby extending the applicability of fixed-Hessian methods beyond specific models such as logistic regression \cite{bohning1988monotonicity,IDASH2018bonte}.
\end{enumerate}

\section{Preliminaries}

In the following discussion, we employ square brackets $[\cdot]$ to denote the indexing of vectors and matrices. Specifically, for a vector $\boldsymbol{v} \in \mathbb{R}^{n}$ and a matrix $\mathbf{M} \in \mathbb{R}^{m \times n}$, $\boldsymbol{v}[i]$ (or $\boldsymbol{v}_i$) denotes the $i$-th element of $\boldsymbol{v}$, while $\mathbf{M}[i][j]$ (or $\mathbf{M}_{i,j}$) refers to the element in the $i$-th row and $j$-th column of $\mathbf{M}$.

\subsection{Fully Homomorphic Encryption}
Fully Homomorphic Encryption (FHE) is a cryptographic primitive that enables an arbitrary number of additions and multiplications directly on ciphertexts. While Gentry proposed the first feasible FHE scheme using a \textit{bootstrapping} operation in 2009 \cite{gentry2009fully}, such schemes remain computationally intensive. Beyond the inherent complexity of FHE, the efficiency of homomorphic computation is significantly influenced by the choice of dataset encoding and the management of plaintext magnitudes \cite{jaschke2016accelerating}. To address the latter, Cheon et~al. \cite{cheon2017homomorphic} introduced an HE scheme featuring a \textit{rescaling} procedure that effectively mitigates this technical bottleneck. 

In our implementation of homomorphic LR algorithms, we adopt the open-source library \texttt{HEAAN}, which realizes this rescaling functionality. Furthermore, leveraging ciphertext packing is essential to achieve a superior amortized runtime. \texttt{HEAAN} supports Single Instruction, Multiple Data (\texttt{SIMD}) operations \cite{smart2014fully}, allowing multiple complex numbers to be packed into the slots of a single ciphertext polynomial and enabling efficient rotation operations across these slots. The mathematical foundations of the \texttt{HEAAN} scheme are detailed in \cite{IDASH2018Andrey, kim2018secure, han2018efficient}, with the underlying abstract algebra further elaborated in \cite{artin2011algebra}.

\subsection{Database Encoding Method} \label{basic he operations}
Kim et~al. \cite{IDASH2018Andrey} proposed a highly efficient database encoding method that leverages \texttt{SIMD} batching to optimize both computational and storage resources. Given a training dataset $\mathbf{Z}$ containing $n$ samples, each with $d+1$ covariates, the authors packed the entire matrix into a single ciphertext in a row-major format, thereby maximizing the utilization of available ciphertext slots.

Under this encoding paradigm, the data matrix $\mathbf{Z}$ can be manipulated directly through homomorphic operations on its ciphertext representation, $\text{Enc}(\mathbf{Z})$. This is achieved using a minimal set of primitive operations: rotation, addition, and multiplication. For example, to isolate the first column of $\text{Enc}(\mathbf{Z})$ while masking others, one can multiply the ciphertext by a constant matrix $\mathbf{Z}$ (where the first column contains ones and all other elements are zeros).

Han et~al. \cite{han2018efficient} delineated several fundamental yet critical operations employed in the implementation by Kim et~al. \cite{IDASH2018Andrey}, such as the \texttt{SumColVec} procedure for computing column-wise summations of a matrix. By composing these primitive operations, more sophisticated computations---most notably the gradient evaluation required for logistic regression---can be efficiently realized in the encrypted domain.

\subsection{Logistic Regression Model}
Logistic regression  is widely used in binary classification tasks to infer whether a binary-valued variable belongs to a certain class or not. LR generalizes linear regression \cite{murphy2012machine} by mapping the linear predictor $\boldsymbol{\beta}^{\top} \mathbf{x}$ to the interval $(0, 1)$ via the sigmoid function $\sigma(z) = (1 + \exp(-z))^{-1}$. Here, $\boldsymbol{\beta} \in \mathbb{R}^{d+1}$ denotes the model parameters, and $\mathbf{x} = (1, x_1, \ldots, x_d)^{\top} \in \mathbb{R}^{d+1}$ represents the input covariate vector. For a class label $y \in \{+1, -1\}$, the conditional probabilities are formulated as:

\begin{equation*}
\begin{aligned}
\Pr(y=+1 \mid \mathbf{x}, \boldsymbol{\beta}) &= \sigma(\boldsymbol{\beta}^{\top} \mathbf{x}) = \frac{1}{1 + e^{-\boldsymbol{\beta}^{\top} \mathbf{x}}}, \\
\Pr(y=-1 \mid \mathbf{x}, \boldsymbol{\beta}) &= 1 - \sigma(\boldsymbol{\beta}^{\top} \mathbf{x}) = \frac{1}{1 + e^{\boldsymbol{\beta}^{\top} \mathbf{x}}}.
\end{aligned}
\end{equation*}

A classification decision is typically made by comparing the output probability against a predefined threshold (e.g., $0.5$).

Training an LR model is framed as a Maximum Likelihood Estimation (MLE) problem, seeking $\boldsymbol{\beta}$ that maximizes the likelihood $L(\boldsymbol{\beta}) = \prod_{i=1}^{n} \Pr(y_i \mid \mathbf{x}_i, \boldsymbol{\beta})$. For computational tractability, we maximize the log-likelihood function $l(\boldsymbol{\beta})$:

\begin{equation*}
l(\boldsymbol{\beta}) = \ln L(\boldsymbol{\beta}) = -\sum_{i=1}^{n} \ln (1 + e^{-y_i \boldsymbol{\beta}^{\top} \mathbf{x}_i}),
\end{equation*}
where $n$ denotes the number of training samples. Since $l(\boldsymbol{\beta})$ lacks a closed-form solution, parameters are typically estimated using iterative optimization: (a) first-order methods, such as gradient descent; and (b) second-order methods, such as Newton’s method. The gradient and Hessian of $l(\boldsymbol{\beta})$ are given by:

\begin{equation*}
\begin{aligned}
\nabla_{\boldsymbol{\beta}} l(\boldsymbol{\beta}) &= \sum_{i=1}^n (1 - \sigma(y_i \boldsymbol{\beta}^{\top} \mathbf{x}_i)) y_i \mathbf{x}_i, \\
\nabla_{\boldsymbol{\beta}}^2 l(\boldsymbol{\beta}) &= \sum_{i=1}^n (y_i \mathbf{x}_i) \sigma(y_i \boldsymbol{\beta}^{\top} \mathbf{x}_i) (\sigma(y_i \boldsymbol{\beta}^{\top} \mathbf{x}_i) - 1) (y_i \mathbf{x}_i)^{\top} = -\mathbf{X}^{\top} \mathbf{S} \mathbf{X},
\end{aligned}
\end{equation*}
where $\mathbf{S}$ is a diagonal matrix with entries $\mathbf{S}_{ii} = \sigma(y_i \boldsymbol{\beta}^{\top} \mathbf{x}_i)(1 - \sigma(y_i \boldsymbol{\beta}^{\top} \mathbf{x}_i))$, and $\mathbf{X} \in \mathbb{R}^{n \times (d+1)}$ denotes the dataset matrix.

The log-likelihood $l(\boldsymbol{\beta})$ is a concave function with at most one unique global maximum \cite{Allison2008LRConvergenceFail}, where its gradient is zero. Newton's method, as a second-order technique, finds the optimal $\boldsymbol{\beta}$ by iteratively solving for the roots of the optimality condition $\nabla_{\boldsymbol{\beta}} l(\boldsymbol{\beta}) = 0$.

\section{Technical Details}
Newton's method requires recomputing the Hessian matrix and its inverse at each iteration, a process that is often computationally prohibitive. To mitigate this burden, the \textit{fixed Hessian Newton method} (also known as the \textit{lower bound principle}) replaces the varying Hessian with a constant matrix $\bar{\mathbf{H}}$, thereby enhancing computational efficiency. 

B\"ohning and Lindsay \cite{bohning1988monotonicity} established that the convergence of Newton's method is guaranteed provided that $\bar{\mathbf{H}} \preceq \nabla_{\boldsymbol{\beta}}^2 l(\boldsymbol{\beta})$, where $\bar{\mathbf{H}}$ is a symmetric negative-definite matrix independent of $\boldsymbol{\beta}$. Here, ``$\preceq$'' denotes the Loewner partial order, implying that the difference $\nabla_{\boldsymbol{\beta}}^2 l(\boldsymbol{\beta}) - \bar{\mathbf{H}}$ is positive semi-definite. By employing such a fixed Hessian $\bar{\mathbf{H}}$, the Newton iteration simplifies to the following update rule:
\begin{equation*}
    \boldsymbol{\beta}_{t+1} = \boldsymbol{\beta}_{t} - \bar{\mathbf{H}}^{-1} \nabla_{\boldsymbol{\beta}} l(\boldsymbol{\beta}_t).
\end{equation*}
B\"ohning and Lindsay further suggest that the fixed matrix $\bar{\mathbf{H}} = -\frac{1}{4}\mathbf{X}^{\top}\mathbf{X}$ serves as an effective lower bound for the Hessian of the log-likelihood function $l(\boldsymbol{\beta})$ in logistic regression.

\subsection{Simplified Fixed Hessian}
Bonte and Vercauteren \cite{IDASH2018bonte} further simplify the fixed Hessian $\bar{\mathbf{H}}$ to facilitate its inversion within the encrypted domain. Specifically, they approximate $\bar{\mathbf{H}}$ with a diagonal matrix $\mathbf{B}$, where each diagonal element $b_{ii}$ is defined as the sum of the corresponding row in $\bar{\mathbf{H}}$. Furthermore, they propose an optimized computational schedule to construct $\mathbf{B}$ more efficiently. This diagonal approximation is formulated as follows:
$$
\mathbf{B} =
\begin{bmatrix}
\sum_{j=0}^{d} \bar{h}_{0j} & 0 & \ldots & 0 \\
0 & \sum_{j=0}^{d} \bar{h}_{1j} & \ldots & 0 \\
\vdots & \vdots & \ddots & \vdots \\
0 & 0 & \ldots & \sum_{j=0}^{d} \bar{h}_{dj}
\end{bmatrix},
$$
where $\bar{h}_{kj}$ denotes the $(k, j)$-th entry of $\bar{\mathbf{H}}$. 

This diagonal structure significantly simplifies the matrix representation, allowing $\mathbf{B}$ to be efficiently derived from $\bar{\mathbf{H}}$. The inverse $\mathbf{B}^{-1}$ can be homomorphically approximated by computing the reciprocal of each diagonal element via an iterative Newton-Raphson method with a carefully chosen initial value. Consequently, the SFH update rule is formulated as:
\begin{equation*}
\begin{aligned}
\boldsymbol{\beta}_{t+1} &= \boldsymbol{\beta}_{t} - \mathbf{B}^{-1} \nabla_{\boldsymbol{\beta}} l(\boldsymbol{\beta}_t) \\
&= \boldsymbol{\beta}_{t} - 
\begin{bmatrix}
b_{00}^{-1} & 0 & \ldots & 0 \\
0 & b_{11}^{-1} & \ldots & 0 \\
\vdots & \vdots & \ddots & \vdots \\
0 & 0 & \ldots & b_{dd}^{-1}
\end{bmatrix} 
\begin{bmatrix}
\nabla_0 \\
\nabla_1 \\
\vdots \\
\nabla_d
\end{bmatrix} \\
&= \boldsymbol{\beta}_{t} - 
\begin{bmatrix}
b_{00}^{-1} \cdot \nabla_0 \\
b_{11}^{-1} \cdot \nabla_1 \\
\vdots \\
b_{dd}^{-1} \cdot \nabla_d
\end{bmatrix},
\end{aligned}
\end{equation*}
where $b_{ii}^{-1}$ denotes the reciprocal of $\sum_{j=0}^{d} \bar{h}_{ij}$, and $\nabla_i$ represents the $i$-th element of the gradient vector $\nabla_{\boldsymbol{\beta}} l(\boldsymbol{\beta}_t)$.

Consider a specific scenario where all diagonal elements $b_{ii}^{-1}$ share a constant value $-\eta$ (with $\eta > 0$). In this case, the iterative update for the SFH method reduces to:
\begin{equation*}
\boldsymbol{\beta}_{t+1} = \boldsymbol{\beta}_{t} - (-\eta) 
\begin{bmatrix}
\nabla_0 \\
\nabla_1 \\
\vdots \\
\nabla_d
\end{bmatrix} = \boldsymbol{\beta}_{t} + \eta \nabla_{\boldsymbol{\beta}} l(\boldsymbol{\beta}_t),
\end{equation*}
which is mathematically equivalent to the vanilla gradient \textit{ascent} update. This alignment provides the foundational insights for the unified framework proposed in this study, which reconciles fixed-Hessian methods with gradient-based optimization. 

We characterize $\mathbf{B}^{-1} \nabla_{\boldsymbol{\beta}} l(\boldsymbol{\beta})$ as an \textit{augmented gradient} and incorporate a dedicated learning rate to capture its underlying dynamics. Provided this learning rate decays from a value exceeding unity (e.g., $2.0$) toward $1.0$ within a finite number of iterations, the fixed-Hessian principle theoretically ensures the convergence of our proposed scheme.

Despite its advantages, the SFH method \cite{IDASH2018bonte} suffers from two primary limitations. First, its convergence is strictly guaranteed only when all entries of the symmetric matrix $\bar{\mathbf{H}}$ are non-positive. While feature normalization (e.g., scaling to $[0, 1]$) often satisfies this condition in machine learning, it may not hold in general numerical optimization. Second, both the simplified Hessian $\mathbf{B}$ and the fixed Hessian $\bar{\mathbf{H}} = -\frac{1}{4}\mathbf{X}^{\top}\mathbf{X}$ are prone to singularity, particularly in high-dimensional sparse settings such as the MNIST dataset. 

To address these issues, we generalize the SFH framework to ensure the invertibility of the simplified Hessian under arbitrary conditions. Building on this generalization, we introduce an accelerated gradient variant, termed the \textit{quadratic gradient}.

\subsection{Quadratic Gradient: Definition}
Suppose a differentiable scalar-valued function $F(\mathbf{x})$ has a gradient $\boldsymbol{g}$ and a Hessian matrix $\mathbf{H}$. For the maximization task, let $\bar{\mathbf{H}}$ be any matrix satisfying the condition $\bar{\mathbf{H}} \preceq \mathbf{H}$ in the Loewner partial ordering. We represent these quantities as:
\begin{equation*}
\begin{aligned}
& \boldsymbol{g} = \begin{bmatrix} g_0 \\ g_1 \\ \vdots \\ g_d \end{bmatrix}, \quad
\mathbf{H} = \begin{bmatrix} 
\nabla_{00}^2 & \nabla_{01}^2 & \dots & \nabla_{0d}^2 \\ 
\nabla_{10}^2 & \nabla_{11}^2 & \dots & \nabla_{1d}^2 \\ 
\vdots & \vdots & \ddots & \vdots \\ 
\nabla_{d0}^2 & \nabla_{d1}^2 & \dots & \nabla_{dd}^2 
\end{bmatrix}, \\ 
& \quad \quad \quad 
\bar{\mathbf{H}} = \begin{bmatrix} 
\bar{h}_{00} & \bar{h}_{01} & \dots & \bar{h}_{0d} \\ 
\bar{h}_{10} & \bar{h}_{11} & \dots & \bar{h}_{1d} \\ 
\vdots & \vdots & \ddots & \vdots \\ 
\bar{h}_{d0} & \bar{h}_{d1} & \dots & \bar{h}_{dd} 
\end{bmatrix},
\end{aligned}
\end{equation*}
where $\nabla_{ij}^2 = \frac{\partial^2 F}{\partial x_i \partial x_j}$. We construct a diagonal matrix $\tilde{\mathbf{B}}$, referred to as the \textit{diagonal Hessian approximation}, where each diagonal entry $\tilde{B}_{kk}$ is defined as:
\begin{equation*}
\tilde{B}_{kk} = -\epsilon - \sum_{i=0}^{d} |\bar{h}_{ki}|.
\end{equation*}
Formally, $\tilde{\mathbf{B}}$ is structured as:
\begin{equation*}
\tilde{\mathbf{B}} = 
\begin{bmatrix}
-\epsilon - \sum_{j=0}^{d} |\bar{h}_{0j}| & 0 & \dots & 0 \\
0 & -\epsilon - \sum_{j=0}^{d} |\bar{h}_{1j}| & \dots & 0 \\
\vdots & \vdots & \ddots & \vdots \\
0 & 0 & \dots & -\epsilon - \sum_{j=0}^{d} |\bar{h}_{dj}|
\end{bmatrix},
\end{equation*}
where $\epsilon > 0$ is a small constant (typically set to $1\text{e}{-8}$) introduced for numerical stability, such as ensuring strict negative definiteness and preventing division by zero during inversion.

For the approximation $\tilde{\mathbf{B}}$ to serve as a valid lower bound within the fixed-Hessian framework, it must satisfy the convergence criterion $\tilde{\mathbf{B}} \preceq \mathbf{H}$. Given our foundational assumption that $\bar{\mathbf{H}} \preceq \mathbf{H}$, this requirement is satisfied by establishing that $\tilde{\mathbf{B}} \preceq \bar{\mathbf{H}}$. Following the analytical approach detailed in \cite{IDASH2018bonte}, it can be shown that $\tilde{\mathbf{B}}$ consistently maintains this lower-bound property relative to $\bar{\mathbf{H}}$, thereby ensuring stable convergence in the encrypted domain without the computational overhead of the full Hessian.

\begin{lemma}
Let $\mathbf{A} \in \mathbb{R}^{n \times n}$ be a symmetric matrix, and let $\mathbf{B}$ be a diagonal matrix with entries $B_{kk} = -\epsilon - \sum_{i=1}^n |A_{ki}|$ for $k = 1, \dots, n$. Then, $\mathbf{B} \preceq \mathbf{A}$.
\end{lemma}

\begin{proof}
By the definition of Loewner ordering, it suffices to prove that the difference matrix $\mathbf{C} = \mathbf{A} - \mathbf{B}$ is positive definite ($\mathbf{C} \succ 0$). The entries of $\mathbf{C}$ are given by $C_{ii} = A_{ii} + \epsilon + \sum_{k=1}^n |A_{ik}|$ for $i = j$, and $C_{ij} = A_{ij}$ for $i \neq j$. According to Gerschgorin’s Circle Theorem, any eigenvalue $\lambda$ of $\mathbf{C}$ lies within at least one disk $|\lambda - C_{ii}| \le \sum_{j \neq i} |C_{ij}|$. Thus,
\begin{equation*}
\lambda \ge C_{ii} - \sum_{j \neq i} |A_{ij}| = A_{ii} + \epsilon + |A_{ii}| + \sum_{j \neq i} |A_{ij}| - \sum_{j \neq i} |A_{ij}| = A_{ii} + |A_{ii}| + \epsilon.
\end{equation*}
Since $A_{ii} + |A_{ii}| \ge 0$, we conclude that $\lambda \ge \epsilon > 0$ for all eigenvalues $\lambda$. Thus, $\mathbf{C}$ is positive definite, implying $\mathbf{B} \preceq \mathbf{A}$.
\end{proof}

\begin{definition}[Quadratic Gradient]
Utilizing the diagonal approximation $\tilde{\mathbf{B}}$, we define the \textit{quadratic gradient} as $\mathbf{G} = \bar{\mathbf{B}} \boldsymbol{g}$, where $\bar{\mathbf{B}}$ is a diagonal matrix with entries $\bar{B}_{kk} = 1 / |\tilde{B}_{kk}|$. For the maximization of $F(\mathbf{x})$, we employ the iterative update rule:
\begin{equation*}
\mathbf{x}_{t+1} = \mathbf{x}_t + N_t \cdot \mathbf{G},
\end{equation*}
where $N_t \ge 1$ is a dynamic learning rate designed to decay toward $1.0$ over a finite number of iterations.
\end{definition}

Minimizing $F(\mathbf{x})$ is equivalent to maximizing $-F(\mathbf{x})$. In this context, $\tilde{\mathbf{B}}$ can be constructed using either a lower bound of the Hessian of $-F(\mathbf{x})$ or an upper bound of the Hessian of $F(\mathbf{x})$. Notably, $\bar{\mathbf{H}}$ may be set to the Hessian $\mathbf{H}$ itself when available, bypassing the need for separate bound estimation.

\subsubsection{Simplified Application-Oriented Version}
While Newton's method is fundamentally a root-finding algorithm for the optimality condition $\nabla F(\mathbf{x}) = 0$, its application depends on the specific optimization objective. In \textit{minimization} scenarios, the Hessian $\mathbf{H}$ is typically positive definite, directing the update toward a local minimum. Conversely, for the \textit{maximization} tasks addressed in this work, $\mathbf{H}$ is assumed to be negative definite.

When utilized in its standalone form, the quadratic gradient $\mathbf{G}$ functions as a \textit{fixed-Hessian Newton update}. This provides a computationally efficient representation that unifies the convergence properties of second-order methods with the structural simplicity of gradient-based updates:
\begin{equation*}
\mathbf{x}_{t+1} = \mathbf{x}_t \pm N_t \cdot \mathbf{G},
\end{equation*}
where the sign is set to $+$ for maximization and $-$ for minimization.

In practical applications where the Hessian matrix $\mathbf{H}$ is explicitly obtainable, we can bypass the formal construction of the quadratic gradient and derive it directly as: 
\begin{subequations} \label{eq:G_definition}
\begin{align}
 \mathbf{G} &= \text{diag}(\bar{B}_{00}, \bar{B}_{11}, \dots, \bar{B}_{dd}) \cdot \boldsymbol{g} \notag \\
 &= \begin{bmatrix}
 \bar{B}_{00} & 0 & \dots & 0 \\
 0 & \bar{B}_{11} & \dots & 0 \\
 \vdots & \vdots & \ddots & \vdots \\
 0 & 0 & \dots & \bar{B}_{dd}
 \end{bmatrix} \cdot \boldsymbol{g} \notag \\
 &= \begin{bmatrix}
 \bar{B}_{00} \\
 \bar{B}_{11} \\
 \vdots \\
 \bar{B}_{dd} 
 \end{bmatrix} \odot \boldsymbol{g}, \label{eq:hadamard_step}
\end{align}
\end{subequations}
where each diagonal element $\bar{B}_{ii}$ is defined as:
\begin{equation*}
\bar{B}_{ii} = \left( \epsilon + \sum_{j=0}^{d} | \nabla_{ij}^2 | \right)^{-1}.
\end{equation*}
Here, $\nabla_{ij}^2 = \frac{\partial^2 F(\mathbf{x})}{\partial x_i \partial x_j}$ denotes the $(i, j)$-th entry of \textit{the Hessian matrix itself}, representing the second-order partial derivative of the objective function with respect to variables $x_i$ and $x_j$. 
The operator ``$\odot$'' denotes the Hadamard product (i.e., element-wise multiplication). This reformulation transitions the computation from a standard matrix-vector product to a vectorized form that is highly compatible with \text{HE-SIMD} parallel processing. Specifically, by transforming the diagonal matrix $\text{diag}(\bar{B}_{00}, \dots, \bar{B}_{dd})$ into a compact vector and computing its Hadamard product with the encrypted vector $\boldsymbol{g}$, the cloud server can execute the entire operation in a single homomorphic multiplication cycle. This approach optimizes both storage and execution time by circumventing the processing of off-diagonal zero elements.

In such cases, the surrogate Hessian matrix constructed still adheres to the convergence criteria of the fixed-Hessian Newton method. Consistent with our unified framework, the update rule employs an addition ($+$) sign for maximization tasks and a subtraction ($-$) sign for minimization tasks:
\begin{equation*}
\mathbf{x}_{t+1} = \mathbf{x}_t \pm N_t \cdot \mathbf{G},
\end{equation*}
where the choice of sign ensures alignment with the respective optimization objective.

Although Newton’s method is fundamentally a root-finding algorithm, its optimization iteration always follows a subtraction-based update rule, regardless of whether the objective is maximization or minimization. In contrast, the naive quadratic-gradient method can be viewed as a fixed-Hessian Newton variant that replaces the Hessian with a positive-definite surrogate matrix, thereby avoiding Hessian definiteness analysis and naturally inheriting the ascent/descent direction from the corresponding first-order optimizer.


\subsubsection{Systematic Fixed-Hessian Identification}
B\"ohning and Lindsay \cite{bohning1988monotonicity} did not provide a systematic methodology for constructing constant Hessian approximations, potentially due to the scarcity of such matrices for general objective functions. In this work, we introduce a structured procedure to identify and construct these fixed Hessian matrices. Our proposed approach serves as a diagnostic tool: identifying a fixed Hessian via this method provides sufficient evidence of its existence. However, it is important to note that our technique constitutes a \textit{sufficient but not necessary} condition; thus, the inability to detect a fixed Hessian through our procedure does not preclude its existence.

Notably, our quadratic gradient approach does not strictly depend on a $constant (fixed)$ Hessian replacement. Instead, it allows for the use of the original, varying Hessian to construct the dynamic diagonal matrix $\tilde{\mathbf{B}}$. Under this more flexible framework, we present a methodical three-step approach to evaluate the feasibility of employing a fixed Hessian for computational efficiency:

\begin{itemize}
    \item \textbf{Step 1: Construction.} Derive the diagonal Hessian approximation $\tilde{\mathbf{B}}$ directly from the Hessian of the objective function.
  
    \item \textbf{Step 2: Optimization and Bound Identification.} For each diagonal element $\tilde{B}_{kk}$, determine whether a constant maximum for $|\tilde{B}_{kk}|$ exists (equivalent to a constant minimum for $\bar{B}_{kk} = 1 / |\tilde{B}_{kk}|$). If each $|\tilde{B}_{kk}|$ achieves a global upper bound $|\tilde{M}_{kk}|$, then a constant diagonal matrix with entries $1 / |\tilde{M}_{kk}|$ satisfies the convergence criteria for the SFH method.
  
    \item \textbf{Step 3: Performance Evaluation.} Assess the practical performance of the resulting fixed Hessian. While such a replacement ensures theoretical convergence, it may not provide a tight bound to the true Hessian, potentially leading to a suboptimal convergence rate in practice.
\end{itemize}

This systematic search can be applied to re-derive the fixed Hessian matrix $-\frac{1}{4}\mathbf{X}^{\top}\mathbf{X}$ originally proposed in \cite{bohning1988monotonicity} for binary logistic regression, thereby unifying our framework with classical results.

From another perspective, the standard first-order gradient update, $\boldsymbol{\beta}_{t+1} = \boldsymbol{\beta}_{t} - \alpha \boldsymbol{g}$ (where $\alpha$ denotes a sufficiently small step size), can be interpreted as a degenerate instantiation of the fixed-Hessian Newton method. This equivalence holds under the condition that, with high probability, the scalar $\alpha$ is stochastically bounded by the diagonal elements of the Hessian approximation, i.e., $\alpha \leq \bar{B}_{kk}$ for all $k \in \{1, \dots, d\}$, where $\bar{B}_{kk}$ represents the $k$-th diagonal entry defined in the preceding Step 2. Consequently, the gradient descent update effectively serves as an isotropic approximation of second-order curvature, where the learning rate $\alpha$ implicitly regularizes the spectral properties of the approximated Hessian $\bar{\mathbf{B}}$. In this view, the inverse of the surrogate Hessian is simplified to a uniform scalar $\alpha \mathbf{I}$. However, such a scalar approximation fails to capture the non-uniform curvature of the objective function. Consequently, while it offers baseline stability, it lacks the adaptive convergence required for practical efficiency—particularly in high-dimensional encrypted optimization, where the high computational cost of each iteration necessitates a minimal number of steps to reach convergence.



\subsubsection{Unitary Scalars as Degenerate Vectors}
In numerical optimization, acceleration techniques for Newton-type methods primarily aim to enhance convergence robustness while mitigating computational overhead. A prevalent strategy is \textit{damping} (or line search), where the full Newton step is modulated by a scaling factor to ensure global convergence. Mathematically, this damping parameter acts as a scalar multiplier, functionally analogous to the \textit{learning rate} in first-order methods. However, unlike fixed step sizes, this parameter is adaptively determined to balance aggressive local progress with global stability, serving as a curvature-aware step controller.

From a structural perspective, the scalar parameter $\alpha$ in line-search methods can be interpreted as a \textit{degenerate vector} where all entries are identical, i.e., $[\alpha, \alpha, \dots, \alpha]^\top$. This representation reveals an inherent informational redundancy, where a scalar is essentially a high-dimensional vector with perfectly correlated components. In both first-order and second-order optimization, a standard learning rate or a line-search factor $\alpha$ imposes a uniform scaling across all dimensions of the update space. Under this interpretation, a scalar step size is not fundamentally distinct from a vector; rather, it represents a specific homogeneous structure where the update magnitude is constrained to be identical across all degrees of freedom.

We contend that the search space for Newton-type methods should ideally be represented by a search matrix $\mathbf{S} \in \mathbb{R}^{d \times d}$, which is isomorphic to the Hessian matrix $\mathbf{H}$. The general update rule is formulated as:
\begin{equation*}
\boldsymbol{\beta}_{t+1} = \boldsymbol{\beta}_{t} - \mathbf{S} \mathbf{H}^{-1} \nabla_{\boldsymbol{\beta}} l(\boldsymbol{\beta}_t).
\end{equation*}
Standard line-search techniques drastically simplify this search space by substituting the full matrix $\mathbf{S}$ with a scalar $\alpha$. Consequently, this scalar acts as a degenerate identity matrix $\alpha \mathbf{I}$, collapsing the complexity of the directional update from $\mathcal{O}(d^2)$ to $\mathcal{O}(1)$ and constraining the optimization trajectory to a single-degree-of-freedom manifold. 

By expanding this collapsed scalar into a non-homogeneous vector, our proposed quadratic gradient framework restores the essential degrees of freedom required for efficient convergence in high-dimensional landscapes, agnostic of whether the optimization is conducted in the encrypted or plaintext domain. Consequently, we contend that Newton acceleration techniques relying on a single scalar, such as conventional line search, represent a degenerate special case of our quadratic gradient framework.

This evolutionary trajectory—moving from constrained scalar updates toward high-dimensional adaptive control—mirrors the historical development of first-order optimization: progressing from fixed step sizes (\textit{Vanilla} SGD) to adaptive scalar rates (NAG), and ultimately to element-wise adaptive algorithms (e.g., AdaGrad, Adam). In the following sections, we demonstrate that the learning rate mechanism within our quadratic gradient framework undergoes a parallel transition. For instance, our \textit{Enhanced Adam} algorithm can be viewed as a formal integration of the fixed-Hessian Newton method with a modernized, element-wise adaptive learning scheme.

\begin{algorithm}[htbp]
    \caption{The Enhanced Nesterov's Accelerated Gradient Algorithm\protect\footnotemark}
    \label{alg:enhanced_nag_algorithm}
    \begin{algorithmic}[1]
        \STATE {\bfseries Require:} training dataset $X \in \mathbb{R}^{n \times (1+d)}$; training label $Y \in \mathbb{R}^{n \times 1}$; learning rate $lr \in \mathbb{R}$; and the number $\kappa$ of iterations;
        \STATE {\bfseries Ensure:} the parameter vector $V \in \mathbb{R}^{(1+d)}$

        \STATE {\bfseries Set} $\bar H \gets -\frac{1}{4}X^{\top}X$
        \# $\bar H \in \mathbb{R}^{(1+d) \times (1+d)}$

        \STATE {\bfseries Set} $V \gets \boldsymbol 0$, $W \gets \boldsymbol 0$, $\bar B \gets \boldsymbol 0$
        \# $V \in \mathbb{R}^{(1+d)}$, $W \in \mathbb{R}^{(1+d)}$, $\bar B \in \mathbb{R}^{(1+d) \times (1+d)}$

        \FOR{$i := 0$ to $d$}
            \STATE \# $\epsilon$ is a small positive constant such as $1e-8$            
            \STATE $\bar B[i][i] \gets \epsilon$         
            \FOR{$j := 0$ to $d$}
                \STATE $\bar B[i][i] \gets \bar B[i][i] + |\bar H[i][j]|$
            \ENDFOR
        \ENDFOR

        \STATE {\bfseries Set} $alpha_0 \gets 0.01$
        \STATE {\bfseries Set} $alpha_1 \gets 0.5 \times (1 + \sqrt{1 + 4 \times alpha_0^2})$

        \FOR{$count := 1$ to $\kappa$}
            \STATE \# $Z \in \mathbb{R}^{n}$  will store  the inputs for  Sigmoid function            
            \STATE {\bfseries Set} $Z \gets \boldsymbol 0$
            \FOR{$i := 1$ to $n$}
                \FOR{$j := 0$ to $d$}
                    \STATE $Z[i] \gets Z[i] + Y[i] \times V[j] \times X[i][j]$
                \ENDFOR
            \ENDFOR
      
      \STATE \# $\boldsymbol \sigma \in \mathbb{R}^{n}$ will store the outputs of Sigmoid function      
            \STATE {\bfseries Set} $\boldsymbol \sigma \gets \boldsymbol 0$
            \FOR{$i := 1$ to $n$}
                \STATE $\boldsymbol \sigma[i] \gets 1 / (1 + \exp(-Z[i]))$
            \ENDFOR

            \STATE {\bfseries Set} $\boldsymbol g \gets \boldsymbol 0$
            \FOR{$j := 0$ to $d$}
                \FOR{$i := 1$ to $n$}
                    \STATE $\boldsymbol g[j] \gets \boldsymbol g[j] + (1 - \boldsymbol \sigma[i]) \times Y[i] \times X[i][j]$
                \ENDFOR
            \ENDFOR

            \STATE {\bfseries Set} $G \gets \boldsymbol 0$
            \FOR{$j := 0$ to $d$}
                \STATE $G[j] \gets \bar B[j][j] \times \boldsymbol g[j]$
            \ENDFOR

            \STATE {\bfseries Set} $\eta \gets (1 - alpha_0) / alpha_1$
            \STATE {\bfseries Set} $\gamma \gets lr / (n \times count)$
            
            \FOR{$j := 0$ to $d$}
                \STATE $w_{temp} \gets V[j] + (1 + \gamma) \times G[j]$
                \STATE $V[j] \gets (1 - \eta) \times w_{temp} + \eta \times W[j]$
                \STATE $W[j] \gets w_{temp}$
            \ENDFOR

            \STATE $alpha_0 \gets alpha_1$
            \STATE $alpha_1 \gets 0.5 \times (1 + \sqrt{1 + 4 \times alpha_0^2})$
        \ENDFOR
        \STATE {\bfseries Return} $V$
    \end{algorithmic}
\end{algorithm}

\footnotetext{The Python implementation of Algorithm~\ref{alg:enhanced_nag_algorithm}, optimized for Google Colab, is available at: \url{https://anonymous.4open.science/r/IDASH2017-245B/Python3Experiments/}.}

\subsection{Quadratic Gradient: Algorithms}

The quadratic gradient framework can be generalized to enhance various first-order optimization algorithms by incorporating second-order curvature information into their respective update rules:

\begin{enumerate}
    \item \textbf{NAG:} NAG is a variant of the momentum method designed to provide the momentum term with greater prescience. The standard iterative formulas for NAG in a gradient \textit{ascent} context are:
    \begin{align}
        \mathbf{V}_{t+1} &= \boldsymbol{\beta}_{t} + \eta_t \nabla J(\boldsymbol{\beta}_t), \label{eq:nag_intermediate} \\
        \boldsymbol{\beta}_{t+1} &= (1-\gamma_t) \mathbf{V}_{t+1} + \gamma_t \mathbf{V}_{t}, \label{eq:nag_final}
    \end{align}
    where $\mathbf{V}_{t+1}$ is an intermediate variable and $\gamma_t \in (0, 1)$ is a smoothing parameter for the moving average, enabling gradient evaluation at an approximate future position \cite{IDASH2018Andrey}. The \textit{Enhanced NAG} algorithm modifies the original update rule \eqref{eq:nag_intermediate} by substituting the vanilla gradient with the quadratic gradient:
    \begin{equation*}
        \mathbf{V}_{t+1} = \boldsymbol{\beta}_{t} + N_t \mathbf{B}^{-1} \nabla J(\boldsymbol{\beta}_t) = \boldsymbol{\beta}_{t} + N_t \mathbf{G}_t,
    \end{equation*}
    where $N_t$ is the adaptive learning rate for the enhanced variant. As detailed in Algorithm~\ref{alg:enhanced_nag_algorithm}, we typically set $N_t = 1 + \eta_t$, where $\eta_t$ follows a specific decay schedule.

    \item \textbf{AdaGrad:} AdaGrad is particularly well-suited for sparse data by adapting the learning rate for each parameter based on historical gradients. The update rules for standard AdaGrad and its \textit{Enhanced Quadratic} version for each parameter $\boldsymbol{\beta}_\{(i)\}$ at iteration $t$ are given by:
    \begin{equation*}
      \begin{aligned}
      \text{(Standard)} \quad \boldsymbol{\beta}_{t+1}^{(i)} &= \boldsymbol{\beta}_t^{(i)} - \frac{\eta_t}{\epsilon + \sqrt{\sum_{k=1}^t (\boldsymbol{g}_k^{(i)})^2}} \cdot \boldsymbol{g}_t^{(i)}, \\
      \text{(Enhanced)} \quad \boldsymbol{\beta}_{t+1}^{(i)} &= \boldsymbol{\beta}_t^{(i)} - \frac{N_t}{\epsilon + \sqrt{\sum_{k=1}^t (\mathbf{G}_k^{(i)})^2}} \cdot \mathbf{G}_t^{(i)}.
     \end{aligned}
    \end{equation*}

    \item \textbf{Adam:} Adam integrates the benefits of AdaGrad's element-wise adaptation with the stability of momentum, enabling faster convergence in the presence of sparse or noisy gradients. Following the same logic as the AdaGrad enhancement, the \textit{Enhanced Adam} algorithm replaces the vanilla gradient with its quadratic form and employs the specialized learning rate within its bias-corrected first and second moment estimates.
\end{enumerate}

\noindent NAG, AdaGrad, and Adam are representative first-order optimization methods that differ primarily in their strategies for learning-rate adaptation and momentum accumulation. The proposed quadratic gradient framework is orthogonal to these mechanisms, as it modifies the gradient representation itself rather than the optimization strategy. Consequently, the same principle can potentially be extended to a broader class of first-order optimization algorithms by replacing the conventional gradient with its quadratic counterpart while preserving the original update structure. 

\begin{figure}[t!]
\centering

\begin{subfigure}{0.23\textwidth}
  \includegraphics[width=\linewidth]{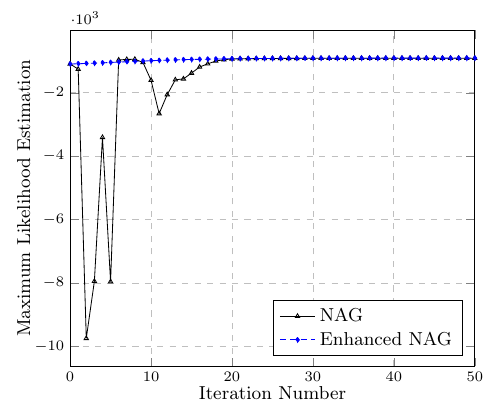}
  \caption{The iDASH dataset}
\end{subfigure}
\hfill
\begin{subfigure}{0.23\textwidth}
  \includegraphics[width=\linewidth]{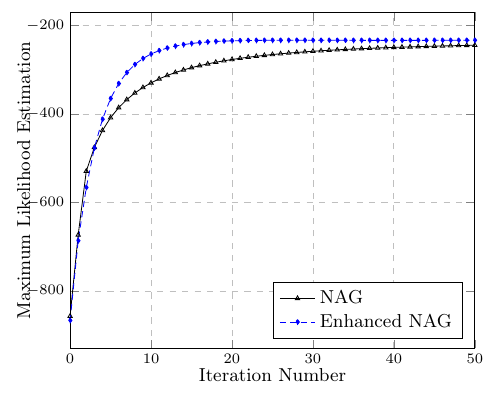}
  \caption{The edin dataset}
\end{subfigure}

\vspace{0.5em}

\begin{subfigure}{0.23\textwidth}
  \includegraphics[width=\linewidth]{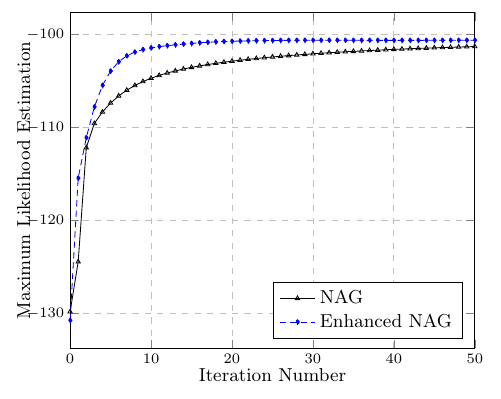}
  \caption{The lbw dataset}
\end{subfigure}
\hfill
\begin{subfigure}{0.23\textwidth}
  \includegraphics[width=\linewidth]{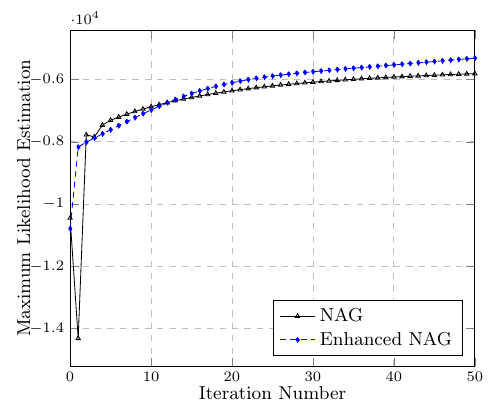}
  \caption{The nhanes3 dataset}
\end{subfigure}

\vspace{0.5em}

\begin{subfigure}{0.23\textwidth}
  \includegraphics[width=\linewidth]{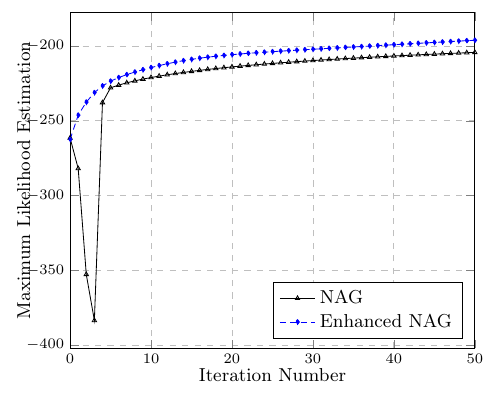}
  \caption{The pcs dataset}
\end{subfigure}
\hfill
\begin{subfigure}{0.23\textwidth}
  \includegraphics[width=\linewidth]{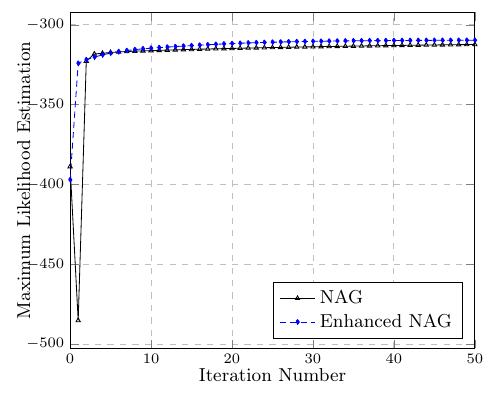}
  \caption{The uis dataset}
\end{subfigure}

\vspace{0.5em}

\begin{subfigure}{0.23\textwidth}
  \includegraphics[width=\linewidth]{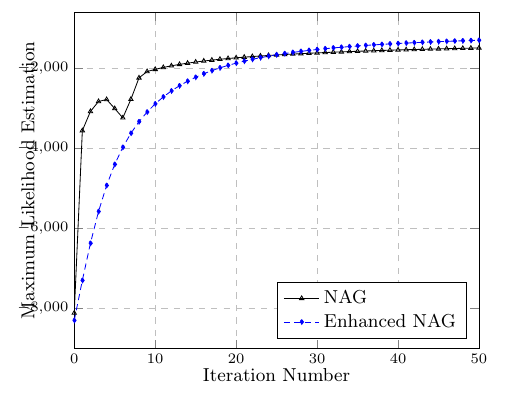}
  \caption{Restructured MNIST dataset}
\end{subfigure}
\hfill
\begin{subfigure}{0.23\textwidth}
  \includegraphics[width=\linewidth]{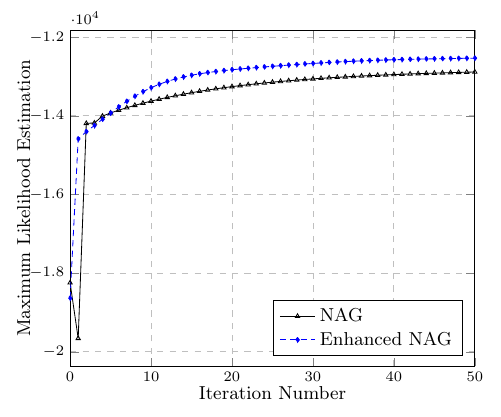}
  \caption{Private financial dataset}
\end{subfigure}

\caption{Training results of NAG and Enhanced NAG in the plaintext domain across eight datasets.}

\Description{
Eight learning-curve plots comparing the standard NAG optimizer and the proposed Enhanced NAG optimizer on the iDASH, edin, lbw, nhanes3, pcs, uis, restructured MNIST, and private financial datasets. Each subplot shows the evolution of model performance during training. Across most datasets, Enhanced NAG converges in fewer iterations and attains equal or better final performance than standard NAG. The relative improvement varies by dataset, reflecting differences in sample size, dimensionality, and optimization difficulty.
}

\label{fig1}
\end{figure}

\begin{figure}[t!]
\centering

\begin{subfigure}{0.23\textwidth} 
  \includegraphics[width=\linewidth]{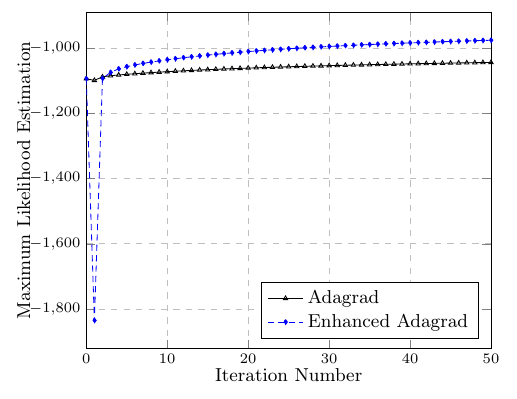}
  \caption{The iDASH dataset}
  \label{fig:subfig01}
\end{subfigure}
\hfill
\begin{subfigure}{0.23\textwidth}
  \includegraphics[width=\linewidth]{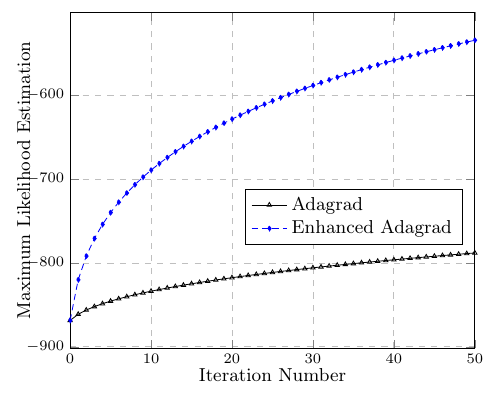}
  \caption{The edin dataset}
\end{subfigure}

\vspace{0.5em}

\begin{subfigure}{0.23\textwidth}
  \includegraphics[width=\linewidth]{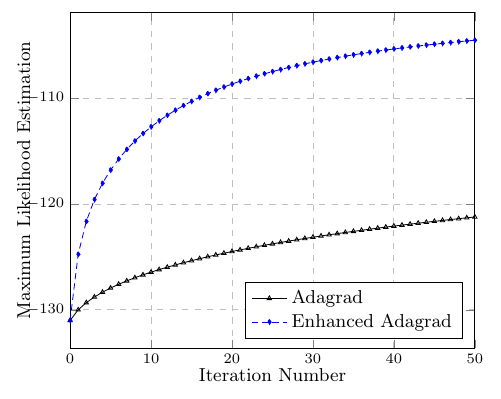}
  \caption{The lbw dataset}
\end{subfigure}
\hfill
\begin{subfigure}{0.23\textwidth}
  \includegraphics[width=\linewidth]{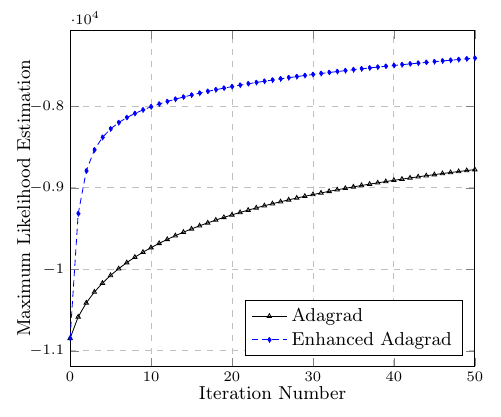}
  \caption{The nhanes3 dataset}
\end{subfigure}

\vspace{0.5em}

\begin{subfigure}{0.23\textwidth}
  \includegraphics[width=\linewidth]{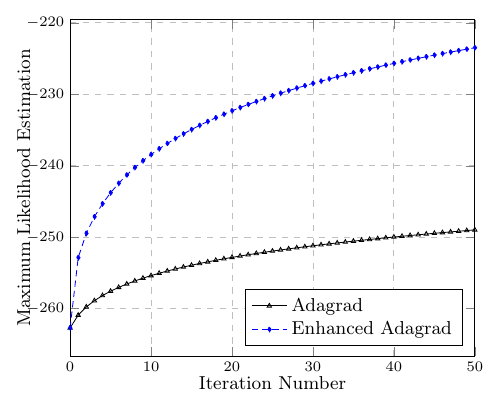}
  \caption{The pcs dataset}
\end{subfigure}
\hfill
\begin{subfigure}{0.23\textwidth}
  \includegraphics[width=\linewidth]{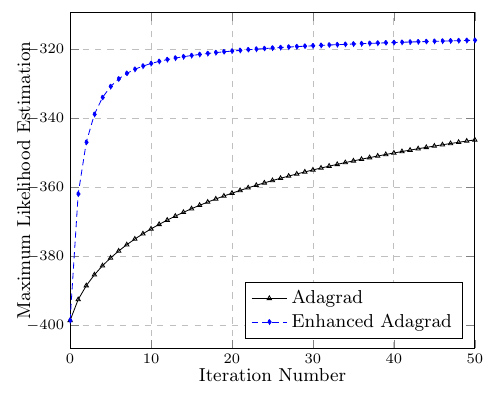}
  \caption{The uis dataset}
\end{subfigure}

\vspace{0.5em}

\begin{subfigure}{0.23\textwidth}
  \includegraphics[width=\linewidth]{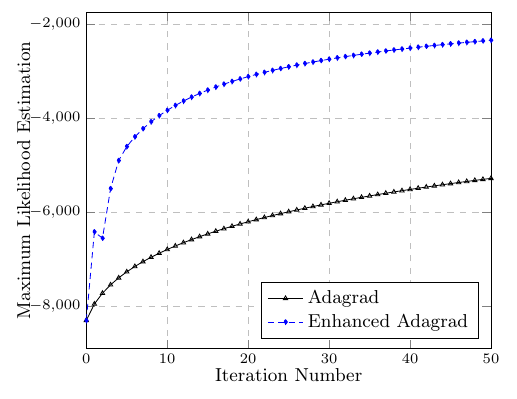}
  \caption{Restructured MNIST dataset}
\end{subfigure}
\hfill
\begin{subfigure}{0.23\textwidth}
  \includegraphics[width=\linewidth]{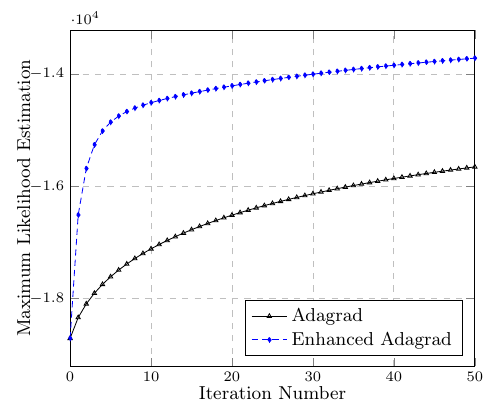}
  \caption{Private financial dataset}
\end{subfigure}

\caption{Training results of Adagrad and Enhanced Adagrad in the plaintext domain across eight datasets.}
\Description{
Eight learning-curve plots comparing Adagrad and Enhanced Adagrad on the iDASH, edin, lbw, nhanes3, pcs, uis, restructured MNIST, and private financial datasets. Each subplot shows model performance over training iterations. Enhanced Adagrad generally converges faster and reaches a higher or comparable final accuracy than the standard Adagrad method. The magnitude of the improvement varies across datasets, with larger gains observed on several medium- and high-dimensional datasets.
}
\label{fig0}

\end{figure}

\begin{figure}[t!]
  \begin{subfigure}{0.23\textwidth}
    \includegraphics[width=\linewidth]{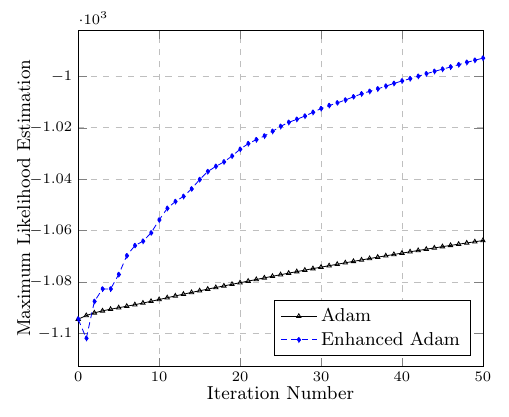}
    \caption{The iDASH dataset}
  \end{subfigure}
  \begin{subfigure}{0.23\textwidth}
    \includegraphics[width=\linewidth]{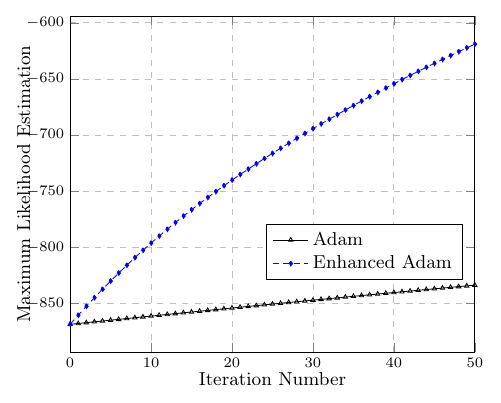}
    \caption{The edin dataset}
  \end{subfigure}\\
  \begin{subfigure}{0.23\textwidth}
    \includegraphics[width=\linewidth]{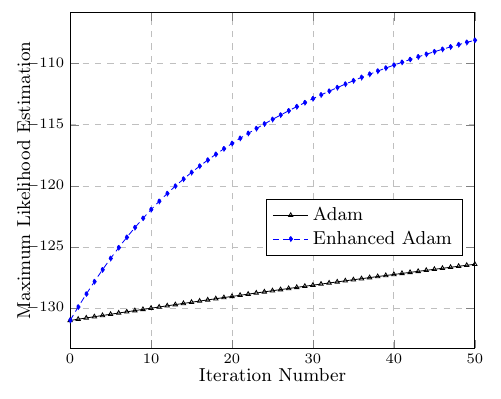}
    \caption{The lbw dataset}
  \end{subfigure}
  \begin{subfigure}{0.23\textwidth}
    \includegraphics[width=\linewidth]{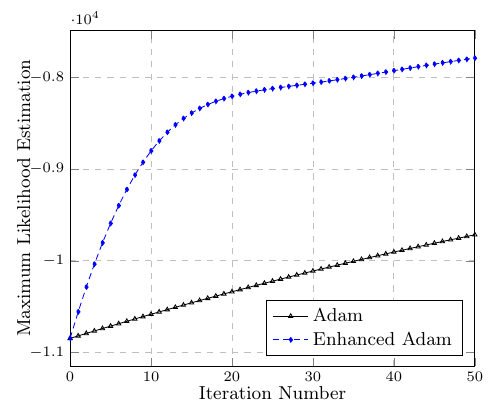}
    \caption{The nhanes3 dataset}
  \end{subfigure}\\
  \begin{subfigure}{0.23\textwidth}
    \includegraphics[width=\linewidth]{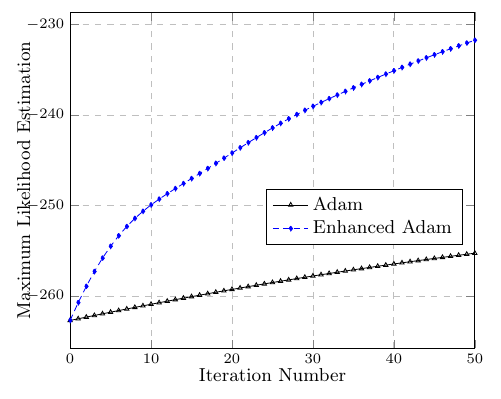}
    \caption{The pcs dataset}
  \end{subfigure}
  \begin{subfigure}{0.23\textwidth}
    \includegraphics[width=\linewidth]{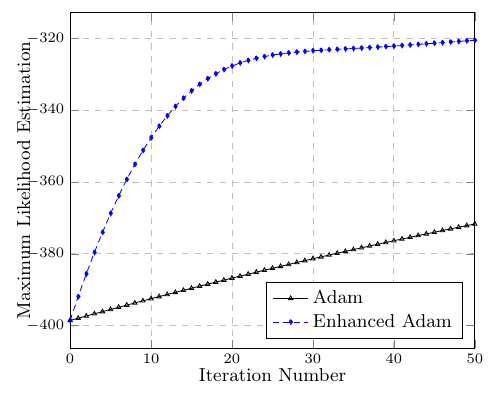}
    \caption{The uis dataset}
  \end{subfigure}\\
  \begin{subfigure}{0.23\textwidth}
    \includegraphics[width=\linewidth]{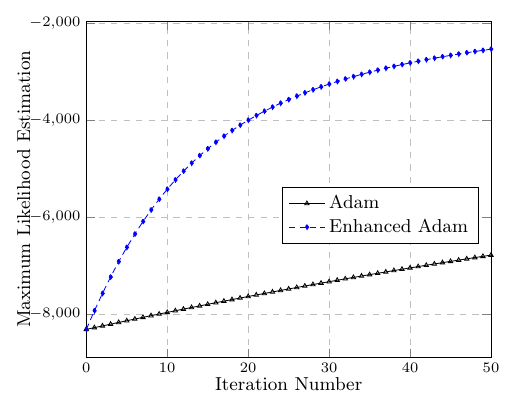}
    \caption{Restructured MNIST dataset}
  \end{subfigure}
  \begin{subfigure}{0.23\textwidth}
    \includegraphics[width=\linewidth]{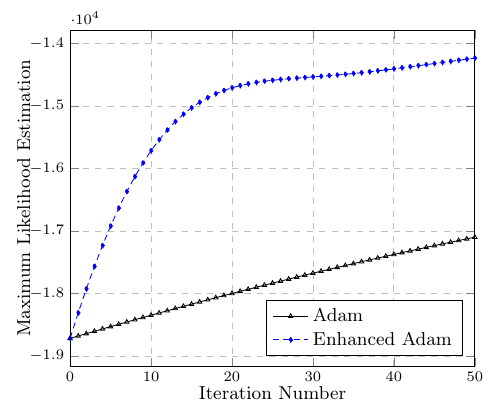}
    \caption{The private financial dataset}
  \end{subfigure}

  \caption{
  Training results of Adam and Enhanced Adam in the plaintext domain.
  }

  \Description{
  Eight subfigures compare the optimization behavior of Adam and Enhanced Adam
  on different datasets: iDASH, edin, lbw, nhanes3, pcs, uis, a
  restructured MNIST dataset, and a private financial dataset. Each subfigure
  presents the training trajectory of the two optimization methods over
  successive iterations. The plots allow comparison of convergence speed,
  stability, and the final objective value achieved by each method. Across the
  datasets, Enhanced Adam generally converges more rapidly and exhibits smoother
  optimization behavior than the standard Adam optimizer.
  }

  \label{fig2}
\end{figure}

\subsubsection{Experiments} \ 

\noindent \textbf{Setup and Datasets}
We evaluate the performance of our proposed algorithms in a non-encrypted context using Python on a workstation equipped with an Intel Core G640 CPU (1.60 GHz) and 7.3 GB of RAM. As our primary objective is to accelerate training convergence, we utilize the log-likelihood function $l(\boldsymbol{\beta})$ under Maximum Likelihood Estimation as the primary performance metric. 

A comprehensive evaluation is conducted on six optimization algorithms: NAG, AdaGrad, Adam, and their respective quadratic-gradient enhanced variants (denoted as Enhanced NAG, Enhanced Adagrad, and Enhanced Adam). We test these on the benchmark datasets utilized by Kim et~al. \cite{IDASH2018Andrey}, which include the iDASH genomic dataset, the Myocardial Infarction dataset (edin), Low Birth Weight Study (lbw), NHANES III (nhanes3), Prostate Cancer Study (pcs), and the Umaru Impact Study (uis). The iDASH dataset, sourced from the 2017 iDASH competition (Task 3), comprises 1,579 records with 103 binary genotypes and a binary phenotype indicating cancer status. 

To assess scalability, we further evaluate these algorithms on two large-scale datasets from \cite{han2018efficient}: a real-world financial dataset containing 422,108 samples and 200 features, and a restructured MNIST dataset featuring 11,982 training samples with 196 features. Throughout our experiments, we consistently employ the fixed Hessian approximation $\bar{\mathbf{H}} = - \frac{1}{4} \mathbf{X}^{\top} \mathbf{X}$ to construct the diagonal surrogate $\tilde{\mathbf{B}}$.



\textbf{Hyperparameter Configuration}
To ensure a fair comparison with the baseline NAG \cite{IDASH2018Andrey}, Enhanced NAG adopts a learning rate of $1 + \frac{10}{1 + t}$, mirroring the decay structure of $\frac{10}{1 + t}$ used in the original work. For Enhanced AdaGrad, we employ a modified learning rate of $N_t = 0.06 + 0.01$, compared to the default $0.01$ used in standard AdaGrad. Similarly, Enhanced Adam adopts an adjusted configuration with a new augmented learning rate of $\alpha = 0.011$ (calculated as the sum of a base rate $0.01$ and the original $0.001$), while maintaining the standard momentum parameters $\beta_1 = 0.9$ and $\beta_2 = 0.999$. In contrast, the baseline Adam algorithm follows its default setting of $\alpha = 0.001$ \cite{kingma2014adam}.

\textbf{Performance Discussion}
As demonstrated in Figures \ref{fig1}--\ref{fig2}, the proposed enhanced algorithms demonstrate both guaranteed convergence and a superior convergence rate compared to their first-order baselines. 

Notably, the learning rates for Enhanced NAG were initially selected to maintain parity with baseline methods rather than for peak optimization. However, the framework demonstrates even more pronounced gains under alternative schedules. For instance, by adopting an exponentially decaying policy ($N_t = 1 + 6 \times 0.9^t$) on the restructured MNIST dataset, Enhanced NAG exhibits a more clearly accelerated convergence trajectory relative to standard NAG.

Overall, the experimental results confirm that the integration of the enhanced gradient framework leads to a systematic improvement in optimization efficiency:
\begin{itemize}
    \item \textbf{Rapid Convergence:} The enhanced variants reach convergence thresholds significantly faster than their original counterparts—an essential attribute for computationally expensive environments like Homomorphic Encryption.
    \item \textbf{Consistent Superiority:} Performance gains are consistent across diverse datasets and high-dimensional architectures, confirming the generalizability of the quadratic gradient.
    \item \textbf{Iteration Efficiency:} Near-optimal results are typically achieved within 4 to 5 iterations, providing a critical advantage in resource-constrained or privacy-preserving settings.
\end{itemize}

Empirically, for the \textit{standalone vanilla quadratic-gradient} method, optimal performance is observed when the learning rate is situated within the interval $[1, 2]$. Beyond a value of $3$, the algorithm typically exhibits divergent behavior. For the \textit{Enhanced NAG} variant, we recommend an \textit{exponentially decaying learning rate} to synchronize rapid early-stage progress with late-stage numerical stability.
\noindent \textbf{Robustness and Generalization.} 
Our proposed algorithm maintains consistent numerical stability and performance invariance even when the input space includes negative values, such as in datasets normalized to the interval $[-1, +1]$. This sign-agnostic property ensures that the framework's efficacy is independent of specific data scaling techniques. 

\begin{figure}[t!]
  \centering

  \begin{subfigure}{0.23\textwidth}
    \includegraphics[width=\textwidth]{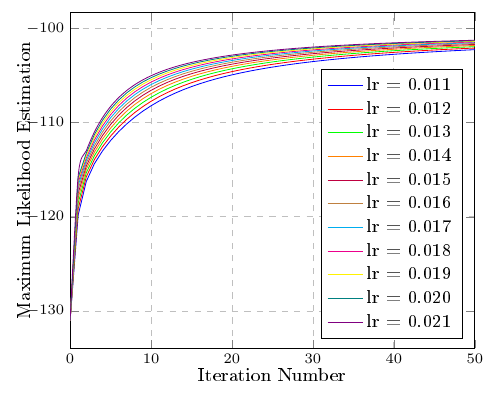}
    \caption{First-order gradient ascent:
    $\boldsymbol{\beta}_{t+1}
    =
    \boldsymbol{\beta}_{t}
    +
    \mathrm{lr}\cdot\boldsymbol{g}$}
    \label{fig:subfig1}
  \end{subfigure}
  \hfill
  \begin{subfigure}{0.23\textwidth}
    \includegraphics[width=\textwidth]{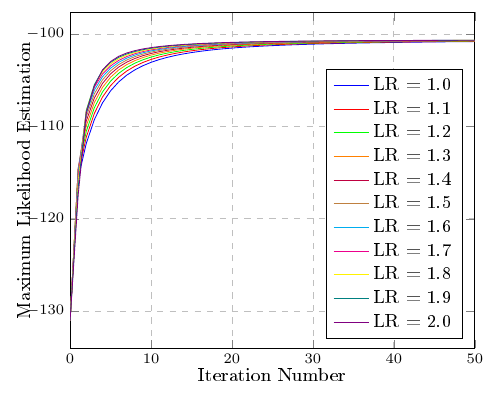}
    \caption{Quadratic gradient ascent:
    $\boldsymbol{\beta}_{t+1}
    =
    \boldsymbol{\beta}_{t}
    +
    \mathrm{LR}\cdot G$}
    \label{fig:subfig2}
  \end{subfigure}

  \caption{
  Training results of standard first-order and quadratic gradient ascent
  algorithms on the \texttt{lbw} dataset in the plaintext domain.
  }

  \Description{
  The figure compares the convergence behavior of two optimization methods on
  the lbw dataset. The left subfigure shows the evolution of the objective
  value obtained by standard first-order gradient ascent using a scalar learning
  rate. The right subfigure shows the corresponding optimization trajectory
  obtained by quadratic gradient ascent using a curvature-aware update matrix.
  The plots illustrate differences in convergence speed, stability, and the
  final objective value achieved by the two methods across training iterations.
  }

  \label{fig3}
\end{figure}

\textbf{Gradient And Quadratic Gradient}
We executed the \textit{first-order vanilla gradient ascent} and the proposed \textit{second-order vanilla quadratic gradient} algorithms on the \textit{lbw} dataset across a spectrum of learning rates, as detailed in Figure~\ref{fig3}. These empirical results provided the foundational inspiration for the conceptualization of the quadratic gradient framework. 

Crucially, our observations reveal that the quadratic gradient inherits the inherent stability and smooth progression characteristics of first-order gradients. Specifically, as the learning rate is incrementally adjusted, both the first-order gradient and second-order quadratic-gradient algorithms exhibit corresponding, continuous shifts in convergence behavior rather than erratic or abrupt fluctuations in performance.

This predictable response to learning rate modulation suggests that the quadratic gradient maintains a well-behaved optimization landscape, effectively bridging the gap between the rapid convergence of second-order methods and the robust, gradual progression of first-order optimization.

\section{Secure Training }
While AdaGrad and Adam offer superior convergence in plaintext, they are often impractical for Homomorphic Encryption due to the high computational cost of frequent inversion and square root operations. Consequently, the \textit{Enhanced NAG} method emerges as the most viable candidate for secure LR training. We leverage this method to implement an efficient, privacy-preserving LR training framework.

A key technical challenge in applying the quadratic gradient is the inversion of the diagonal matrix $\tilde{\mathbf{B}}$ to obtain $\bar{\mathbf{B}}$. To optimize performance, we offload the computation of $\bar{\mathbf{B}}$ to the data owner. Since the data owner prepares and normalizes the dataset, pre-calculating $\bar{\mathbf{B}}$ is both computationally feasible and security-neutral, as it does not leak sensitive information beyond what is already implied by the data normalization process.

\subsection{Polynomial Approximation}
A fundamental constraint in HE-based LR training is the inability of current homomorphic schemes to directly evaluate the non-linear Sigmoid function. To circumvent this, we approximate the Sigmoid function using a least-squares polynomial. Specifically, we utilize the 5th-degree polynomial $g(x)$ developed by Kim et~al. \cite{IDASH2018Andrey}, which fits the Sigmoid function over the interval $[-8, 8]$:
\begin{equation*}
    g(x) = 0.5 + 0.19131 \cdot x - 0.0045963 \cdot x^3 + 0.0000412332 \cdot x^5.
\end{equation*}

\subsection{Data Encoding}
Given a training dataset $\mathbf{X} \in \mathbb{R}^{n \times (1+d)}$ and labels $\mathbf{Y} \in \mathbb{R}^{n \times 1}$, we follow the encoding strategy from \cite{IDASH2018Andrey} to pack the training features and label information into a single ciphertext $\text{ct}_{\mathbf{Z}}$. The initial weight vector $\boldsymbol{\beta}^{(0)}$ (initialized as zeros) and the diagonal entries of $\bar{\mathbf{B}}$ are replicated $n$ times to form matrices suitable for SIMD operations. The data owner then encrypts these into ciphertexts $\text{ct}_{\boldsymbol{\beta}}^{(0)}$ and $\text{ct}_{\bar{\mathbf{B}}}$, formulated as:

\begin{equation*}
 \begin{aligned}
 \mathbf{X} &= 
\begin{bmatrix}
 1  &   x_{11}  &  \dots  &  x_{1d}  \\
 1  &   x_{21}  &  \dots  &  x_{2d}  \\
 \vdots & \vdots & \ddots & \vdots \\
 1  &  x_{n1}   &  \dots  &  x_{nd}  \\
\end{bmatrix}, \quad
\mathbf{Y} = 
\begin{bmatrix}
 y_{1} \\ y_{2} \\ \vdots \\ y_{n}
\end{bmatrix},   \\ 
\text{ct}_{\mathbf{Z}} &= \text{Enc}
\begin{bmatrix}
 y_{1}  &   y_{1}  x_{11} &  \dots  &  y_{1}  x_{1d} \\
 y_{2}  &   y_{2}  x_{21} &  \dots  &  y_{2}  x_{2d} \\
 \vdots & \vdots & \ddots & \vdots \\
 y_{n}  &   y_{n}  x_{n1} &  \dots  &  y_{n}  x_{nd} \\
\end{bmatrix},   
 \end{aligned}
\end{equation*}

\begin{equation*}
 \begin{aligned}
 \text{ct}_{\boldsymbol{\beta}}^{(0)} &= \text{Enc}
\begin{bmatrix}
 \beta_{0}^{(0)}  &   \beta_{1}^{(0)} &  \dots  &  \beta_{d}^{(0)} \\
 \beta_{0}^{(0)}  &   \beta_{1}^{(0)} &  \dots  &  \beta_{d}^{(0)}  \\
 \vdots & \vdots & \ddots & \vdots \\
 \beta_{0}^{(0)}  &   \beta_{1}^{(0)} &  \dots  &  \beta_{d}^{(0)}  \\
\end{bmatrix}, \\
 \text{ct}_{\bar{\mathbf{B}}} &= \text{Enc}
\begin{bmatrix}
  \bar{B}_{00}    & \bar{B}_{11}  &  \dots  & \bar{B}_{dd}  \\
  \bar{B}_{00}    & \bar{B}_{11}  &  \dots  & \bar{B}_{dd}  \\
 \vdots & \vdots & \ddots & \vdots \\
  \bar{B}_{00}    & \bar{B}_{11}  &  \dots  & \bar{B}_{dd}  \\
\end{bmatrix},
 \end{aligned}
\end{equation*}
where $\bar{B}_{ii}$ represents the diagonal entry of $\bar{\mathbf{B}}$ derived from the fixed Hessian $-\frac{1}{4} \mathbf{X}^\top \mathbf{X}$.

The cloud server performs the Enhanced NAG iterations on these ciphertexts to refine the encrypted weight vector. For the detailed homomorphic gradient calculation procedure, we refer the reader to the methodologies in \cite{IDASH2018Andrey}.

\paragraph{Limitations and Trade-offs} 
In a privacy-preserving context, the Enhanced NAG method introduces two primary overheads compared to standard NAG: (1) one additional ciphertext-ciphertext multiplication per iteration to compute the quadratic gradient, and (2) the initial transmission of $\text{ct}_{\bar{\mathbf{B}}}$ by the data owner. However, these costs are effectively compensated by the significantly reduced number of total iterations required to reach convergence, leading to a net reduction in the overall homomorphic execution time. 

Bonte and Vercauteren \cite{IDASH2018bonte} and Ogilvie et al. \cite{ogilvie2020improved} showed that Newton's iterative method with a proper initial approximation can obtain an accurate estimate of $\text{ct}_{\bar{\mathbf{B}}}$ in practical scenarios. Nevertheless, we prefer to delegate this computation to the data owner, since this operation neither leaks sensitive information nor imposes significant computational costs.

\begin{table*}[htbp]
\centering
\caption{Performance comparison on the iDASH dataset using 10-fold cross-validation.}
\label{tab1}

\begin{tabular}{cccccccccc}
\toprule

Dataset &
\mysplit{Sample \\ Num} &
\mysplit{Feature \\ Num} &
Method &
deg $g$ &
\mysplit{Iter \\ Num} &
\mysplit{Storage \\ (GB)} &
\mysplit{Learn \\ Time \\ (min)} &
\mysplit{Accuracy \\ (\%)} &
AUC \\

\midrule

\multirow{2}{*}{iDASH}
& \multirow{2}{*}{1579}
& \multirow{2}{*}{18}
& Ours
& 5
& \cellcolor{lightgray}4
& \cellcolor{lightgray}0.08
& \cellcolor{lightgray}4.43
& \cellcolor{lightgray}61.46
& \cellcolor{lightgray}0.696
\\

&
&
&
\cite{IDASH2018Andrey}
& 5
& 7
& 0.04
& 6.07
& 62.87
& 0.689
\\

\bottomrule

\end{tabular}

\end{table*}

\begin{table*}[htbp]
\centering
\caption{Experimental results on five benchmark datasets under 5-fold cross-validation.}
\label{tab2}

\begin{tabular}{cccccccccc}
\toprule

Dataset &
\mysplit{Sample \\ Num} &
\mysplit{Feature \\ Num} &
Method &
deg $g$ &
\mysplit{Iter \\ Num} &
\mysplit{Storage \\ (GB)} &
\mysplit{Learn \\ Time \\ (min)} &
\mysplit{Accuracy \\ (\%)} &
AUC \\

\midrule

\multirow{2}{*}{edin}
& \multirow{2}{*}{1253}
& \multirow{2}{*}{9}
& Ours
& 5
& \cellcolor{lightgray}4
& \cellcolor{lightgray}0.04
& \cellcolor{lightgray}0.6
& \cellcolor{lightgray}89.52
& \cellcolor{lightgray}0.943
\\

&
&
&
\cite{IDASH2018Andrey}
& 5
& 7
& 0.02
& 3.6
& 91.04
& 0.958
\\

\cmidrule(lr){4-10}

\multirow{2}{*}{lbw}
& \multirow{2}{*}{189}
& \multirow{2}{*}{9}
& Ours
& 5
& \cellcolor{lightgray}4
& \cellcolor{lightgray}0.04
& \cellcolor{lightgray}0.6
& \cellcolor{lightgray}71.35
& \cellcolor{lightgray}0.667
\\

&
&
&
\cite{IDASH2018Andrey}
& 5
& 7
& 0.02
& 3.3
& 69.19
& 0.689
\\

\cmidrule(lr){4-10}

\multirow{2}{*}{nhanes3}
& \multirow{2}{*}{15649}
& \multirow{2}{*}{15}
& Ours
& 5
& \cellcolor{lightgray}4
& \cellcolor{lightgray}0.31
& \cellcolor{lightgray}4.5
& \cellcolor{lightgray}79.23
& \cellcolor{lightgray}0.637
\\

&
&
&
\cite{IDASH2018Andrey}
& 5
& 7
& 0.16
& 7.3
& 79.22
& 0.717
\\

\cmidrule(lr){4-10}

\multirow{2}{*}{pcs}
& \multirow{2}{*}{379}
& \multirow{2}{*}{9}
& Ours
& 5
& \cellcolor{lightgray}4
& \cellcolor{lightgray}0.04
& \cellcolor{lightgray}0.6
& \cellcolor{lightgray}63.20
& \cellcolor{lightgray}0.733
\\

&
&
&
\cite{IDASH2018Andrey}
& 5
& 7
& 0.02
& 3.5
& 68.27
& 0.740
\\

\cmidrule(lr){4-10}

\multirow{2}{*}{uis}
& \multirow{2}{*}{575}
& \multirow{2}{*}{8}
& Ours
& 5
& \cellcolor{lightgray}4
& \cellcolor{lightgray}0.04
& \cellcolor{lightgray}0.6
& \cellcolor{lightgray}74.43
& \cellcolor{lightgray}0.597
\\

&
&
&
\cite{IDASH2018Andrey}
& 5
& 7
& 0.02
& 3.5
& 74.44
& 0.603
\\

\bottomrule
\end{tabular}

\end{table*}

\section{Experiments}
We implement the \textit{Enhanced NAG} algorithm over homomorphically encrypted data using the \texttt{HEAAN} library. For transparency and reproducibility, the C++ source code is publicly available at \href{https://anonymous.4open.science/r/IDASH2017-245B}{https://anonymous.4open.science/r/IDASH2017-245B}. All homomorphic experiments were conducted on a public cloud instance equipped with 32 vCPUs and 64 GB of RAM.

To ensure a fair comparison with the baseline established by Kim et~al. \cite{IDASH2018Andrey}, we adopt an identical 10-fold cross-validation (CV) procedure on the iDASH dataset (1,579 samples, 18 features) and 5-fold CV for the remaining five datasets. We report the average accuracy and Area Under the Curve (AUC) as the primary evaluation metrics. The experimental results, including average runtime and storage consumption (comprising the encrypted dataset for the baseline and the dataset plus $\text{ct}_{\bar{\mathbf{B}}}$ for our method), are summarized in Tables~\ref{tab1} and~\ref{tab2}.

To maintain consistency in ciphertext storage, we utilize the same packing strategy proposed in \cite{IDASH2018Andrey}. This results in comparable ciphertext sizes, with the only additional overhead being the ciphertexts required to encode the diagonal approximation matrix $\bar{\mathbf{B}}$. Throughout the training procedures, we employ a dynamic learning rate of $N_t = 1 + 0.9^t$.

The \texttt{HEAAN} parameters are configured to align closely with the security levels in \cite{IDASH2018Andrey}: $\log N = 16$, $\log Q = 1200$, $\log p = 30$, and the number of slots set to $32,768$, providing a security level of $\lambda \approx 80$. Notably, we utilize a larger scaling factor, $\log p = 40$, to encrypt the matrix $\bar{\mathbf{B}}$ to preserve the precision of the second-order information. 

Due to the increased multiplicative depth—specifically, one additional ciphertext multiplication per iteration compared to the baseline—our scheme consumes the modulus more rapidly. Consequently, without bootstrapping, the evaluation is limited to four iterations of the Enhanced NAG algorithm. Nevertheless, as shown in the results, the proposed method achieves accuracy and AUC levels comparable to the baseline even within this restricted iteration budget. 

Our learning time measurement accounts for these 4 iterations, while the baseline's reported time includes encryption and evaluation. Given that our implementation is built upon their open-source framework, the per-iteration computational cost remains theoretically consistent with the baseline, aside from the marginal overhead of the quadratic gradient multiplication.

\section{Conclusion}
In this paper, we proposed a novel gradient variant, termed the \textit{quadratic gradient}, and developed an \textit{Enhanced NAG} framework specifically tailored for privacy-preserving logistic regression training in the encrypted domain. By directly incorporating second-order curvature information from the Hessian matrix, the proposed quadratic gradient effectively bridges the gap between conventional first-order gradient-based methods and second-order Newton-type optimization techniques. When combined with the Vanilla Quadratic Gradient (namely SFH) or Enhanced NAG, our gradient variants remain compatible with traditional line-search strategies for learning-rate selection. Furthermore, when integrated with adaptive learning rate methods such as NAG, our framework provides a novel methodology for accelerating Newton-type optimization by incorporating embedded gradient-descent steps. This integration preserves the structural simplicity required for homomorphic encryption while achieving improved convergence efficiency.

In the encrypted domain, conventional first-order gradient-based methods, such as NAG, often require careful learning-rate tuning, which partially compromises the non-interactive nature of FHE-based machine learning systems. In contrast, the proposed quadratic gradient framework enables stable optimization with a conservative fixed learning rate. Specifically, both the Vanilla Quadratic Gradient (SFH) and Enhanced NAG can effectively adopt a learning rate of $\eta=1$, thereby eliminating costly learning-rate selection or tuning procedures in encrypted environments. We therefore recommend using a fixed learning rate of 1 for practical deployments of these methods, which alleviates the step-size sensitivity of traditional first-order optimization algorithms and further enhances the practicality of privacy-preserving machine learning.

Several promising directions remain for future research. First, the theoretical foundation of SFH suggests significant potential for accelerating a broader spectrum of numerical optimization algorithms beyond encrypted machine learning applications. In particular, exploring the extension of the quadratic gradient framework to general numerical optimization problems, including convex objective functions, may further demonstrate its effectiveness and theoretical significance.

Second, the applicability of the proposed quadratic gradient framework to non-convex optimization problems remains an important open question. Future studies will investigate its performance in large-scale non-convex settings, with the objective of further validating its scalability, robustness, and generality across diverse optimization landscapes.

Third, although the learning-rate experiments conducted for AdaGrad and Adam in this work successfully identify effective configurations for the evaluated datasets, with the search initialized from $\alpha = 0.1$ for AdaGrad and $\alpha = 0.01$, $\beta_1 = 0.9$, and $\beta_2 = 0.999$ for Adam, the optimal hyperparameter configurations may depend on the characteristics of specific applications and optimization scenarios. Consequently, systematically exploring learning-rate strategies for the proposed enhanced algorithms represents a promising avenue for future research. Specifically, future work may explore whether classical line-search techniques can be efficiently integrated with the Vanilla Quadratic Gradient (SFH) and Enhanced NAG methods, and whether improved learning-rate configurations can further enhance the performance of Enhanced Adam.

Finally, we will investigate the broader applicability of the quadratic gradient framework beyond NAG, AdaGrad, and Adam. Extending this approach to other first-order optimization algorithms may reveal new opportunities for constructing efficient hybrid optimization methods that combine first-order simplicity with second-order curvature awareness.

\appendix

\begin{acks}
The acknowledgments section is defined using the "acks" environment (and NOT an
unnumbered section). This ensures the proper identification of the section in
the article metadata, and the consistent spelling of the heading. Does NOT count
towards page limit. NOT shown (as intended) for anonymous submission. Only edit
for the camera-ready.

Identification of funding sources and other support, and thanks to individuals
and groups that assisted in the research and the preparation of the work should
be included.
\end{acks}

\begin{ethics}
This work focuses on the design and evaluation of privacy-preserving machine learning techniques based on homomorphic encryption. All experiments were conducted using publicly available datasets or synthetically generated data. No human subjects were recruited, no personally identifiable information was collected, and no sensitive user data was accessed.

The purpose of this research is to enhance privacy protections during machine learning training and inference. We are not aware of any direct risks to research participants arising from this work. The techniques developed in this paper are intended to improve data confidentiality in distributed and outsourced computation settings.

Because this research did not involve human participants, human-derived private data, intervention studies, or interaction with individuals, it was not submitted to an Institutional Review Board (IRB) or other external ethics review panel.
\end{ethics}

\begin{openscience}
To support reproducibility, we provide an anonymized repository containing the source code, experimental scripts, benchmark configurations, and documentation used in this work.

The repository has been anonymized to preserve the double-blind review process and is available at:

\url{https://anonymous.4open.science/r/IDASH2017-245B}.

The repository includes instructions for reproducing all experiments and figures reported in the paper.
\end{openscience}

\begin{ai}
The authors used AI-based tools during the preparation of this work. These tools were used to assist with code development, debugging, documentation, literature exploration, and manuscript editing, including improvements to grammar, clarity, and presentation.

All technical designs, algorithms, implementations, experimental results, analyses, and conclusions were developed, verified, and validated by the authors. AI-generated suggestions were reviewed and incorporated only when deemed appropriate.

The authors have manually verified and are responsible for the accuracy, originality, and integrity of the output of all AI-based tools used in this work.
\end{ai}

\bibliographystyle{ACM-Reference-Format}
\bibliography{sample-base}

\end{document}